\documentclass[prl,twocolumn,amsmath]{revtex4-1}
\usepackage{graphicx,epstopdf}
\usepackage{bm}% bold maths
\newcommand{\bracket}[1]{\langle#1\rangle}

\DeclareMathOperator{\RE}{Re}
\DeclareMathOperator{\IM}{Im}
\DeclareMathOperator{\TR}{Tr}

\usepackage[colorlinks=true, letterpaper=true, pdfstartview=FitV,
  linkcolor=blue, citecolor=blue, urlcolor=blue]{hyperref}

\begin{document}

\title{Nonreciprocal Directional Dichroism Induced by the Quantum Metric Dipole}

\author{Yang Gao}

\affiliation{Department of Physics, Carnegie Mellon University,
  Pittsburgh, PA 15213, USA}

\author{Di Xiao}

\affiliation{Department of Physics, Carnegie Mellon University,
  Pittsburgh, PA 15213, USA}

\date{\today}

\begin{abstract}
We identify the quantum metric dipole as the geometric origin of the nonreciprocal directional dichroism which describes the change in the refractive index upon reversing the light propagation direction. Specifically, we find that the static limit of the nonreciprocal directional dichroism corresponds to a quadrupolar transport current from the quantum metric dipole, in response to a quadrupolar electric field. Moreover, at finite frequency, we demonstrate that the steepest slope of the averaged quantum metric dipole determines a peak. Finally, we illustrate both features in a low-energy model.
\end{abstract}

\maketitle

Spatially dispersive optical effects can often yield incisive information on the structural and electronic properties of matter.  A well-known example is natural optical activity in noncentrosymmetric materials~\cite{Landau1984}.  If time-reversal symmetry is also broken, spatial dispersion can give rise to another phenomenon known as nonreciprocal directional dichroism~(NDD)~\cite{Fuchs1965}, referring to the difference in refractive index between counter-propagating lights.  Due to its dependence on both broken time-reversal and inversion symmetry, NDD provides a powerful probe of the dynamical coupling between electricity and magnetism in matter~\cite{Goulon2000, Kubota2004,Arima2008,Kezsmarki2011,Takahashi2011,Okamura2013,Kezsmarki2014,Toyoda2015,Tokura2018}.

To date, the microscopic understanding of NDD has been dominated by molecular theories of electromagnetic multipoles~\cite{Graham1992,Barron2004}. Despite its wide adoption in the literature, this formalism cannot access physics associated with the geometric structure of Bloch states in the momentum space, whose importance has been made increasingly clear in recent years~\cite{Nagaosa2010,Xiao2010,Tokura2018}.  In particular, it has been shown that optical activity and the gyrotropic magnetic effect are connected to the Berry curvature and orbital magnetic moment, revealing their geometric origin~\cite{Zhong2015,Ma2015,Zhong2016}.

In this Letter, we identify the quantum metric as the geometric origin of NDD. Specifically, NDD is connected to the first order moment of quantum metric, which is hence referred to as the quantum metric dipole.  Using the semiclassical transport theory, we show that the integration of the quantum metric dipole over the Fermi surface yields a static current driven by a quadrupolar electric field. Such current is the DC counterpart of NDD and exists in metals with broken time-reversal and inversion symmetry.  Moreover, the quantum metric can be interpreted as the quadrupole moment of the Bloch state.

Our result thus provides an interesting dual to the role of orbital magnetic moment in optical activity: the dependence of the refractive index on light helicity is determined by dipole of orbital magnetic moment~\cite{Ma2015,Zhong2016}, while that on propagation direction is determined by dipole of electric quadrupole moment.

We then use the linear response theory to show that NDD at finite frequency is also determined by the quantum metric dipole. We find that the steepest slope of the averaged quantum metric dipole can give rise to a peak in the differential refractive index between counter-propagating lights.  Finally, we illustrate both the static quadrupolar current and the peak structure of NDD in a low-energy minimal model, which is relevant to van der Waals antiferromagnets.  Our result shows that NDD can be used to probe the geometric structure of Bloch states, and also opens the door to band structure engineering of NDD.

{\it Phenomenological theory of NDD.}---We first give a brief account of the phenomenological theory of NDD in terms of the optical conductivity~\cite{Hornreich1968}.  In medium, the light propagation is characterized by the refractive index $n$, which can be solved from the Maxwell equations. The wave equation for the electric field reads
\begin{equation}\label{eq_maxwell}
\bm \nabla\times (\bm \nabla\times \bm E)=-\mu_0 \frac{\partial \bm J}{\partial t}-\frac{1}{c^2} \frac{\partial^2 \bm E}{\partial t^2}\,,
\end{equation}
where $\mu_0$ is the vacuum permeability and $c$ is the speed of light.  Let us consider a monochromatic light polarized along the $x$-direction and propagating along the $z$-direction. The current $\bm J$ is induced by the electric field through the  conductivity tensor $\sigma_{ij}(\omega,\bm q)$, $J_i(\omega,\bm q)=\sigma_{ij}(\omega,\bm q) E_j(\omega,\bm q)$, with $\omega$ and $\bm q$ being the frequency and wave vector of the light.  If the spatial dispersion is weak, we can expand $\sigma_{ij}(\omega,\bm q)$ in powers of $\bm q$,
\begin{equation}
\sigma_{ij}(\omega,\bm q)=\sigma_{ij}(\omega,0)+\sigma_{ijk}(\omega,0) q_k+\cdots\,.
\end{equation}

The derivation of the refractive index $n$ can be simplified by exerting the following symmetry constraints.  First, we assume the mirror-$z$ symmetry is broken but mirror-$x$ symmetry is present. This forbids the existence of $\sigma_{xy}(\omega,\bm q)$.  We further assume that the system is rotationally invariant about the $z$-axis, so that $\sigma_{xxz}=\sigma_{yyz}$.  By inserting the electric field profile $\bm E\propto e^{i\omega n_zz/c-i\omega t}$ in Eq.~\eqref{eq_maxwell} and using the above symmetry assumptions, we obtain $n_z^2=(n_0+i\kappa_0)^2+ic\mu_0 n_z\sigma_{xxz}$ with $n_0+i\kappa_0=(1+i\mu_0c^2\sigma_{xx}/\omega)^{1/2}$. The solution reads
\begin{equation}\label{eq_nz1}
n_z=ic\mu_0\sigma_{xxz}/2+\sqrt{(n_0+i\kappa_0)^2-c^2\mu_0^2\sigma_{xxz}^2/4}\,.
 \end{equation}
If we reverse the propagation direction, the first term in Eq.~\eqref{eq_nz1} flips sign. Therefore, the difference in the refractive index is $\Delta n=n_z-n_{-z}=ic\mu_0\sigma_{xxz}(\omega,0)$.  Alternatively, we can write~\cite{suppl}
\begin{equation}\label{eq_deltan}
\Delta n=\frac{1}{2}ic\mu_0[\sigma_{xxz}(\omega,0)-\sigma_{xxz}^\star(-\omega^\star,0)]\,.
\end{equation}
Here $\omega\rightarrow \omega+i\eta$ has a small imaginary part.

Note that in the literature the NDD is usually described within the electromagnetic multipole approximation~\cite{Graham1992,Barron2004}.  Here we have taken the Landau-Lifshitz approach by eliminating the magnetic field from our theory via the Maxwell equation.  These two approaches are equivalent~\cite{Bungay1993}.

\begin{figure}[t]
  \includegraphics[width=0.8\columnwidth]{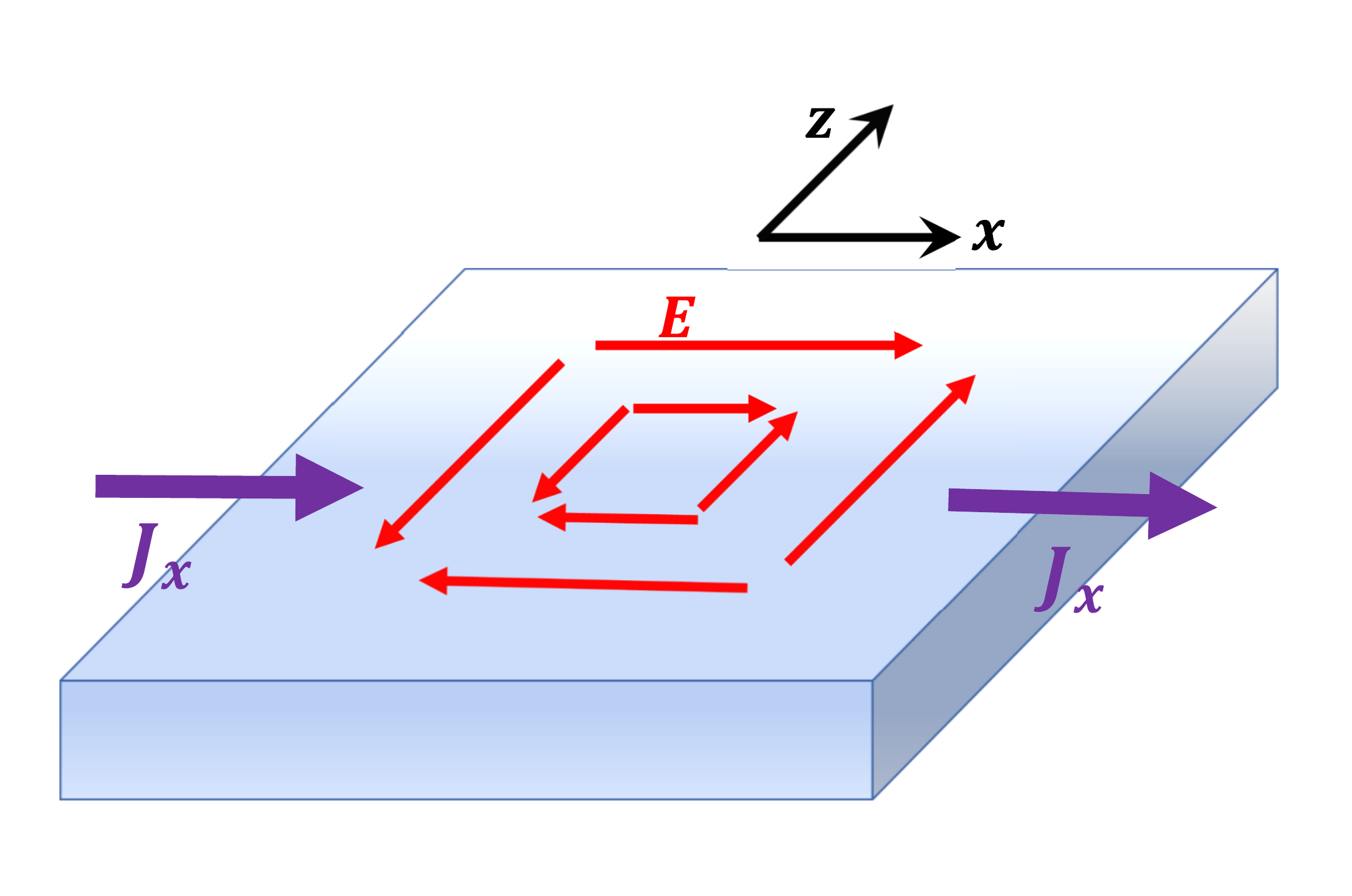}
  \caption{The electric quadrupolar current. The electric field in red arrows has the quadrupolar profile. The resulting quadrupolar current is in purple arrows.}
  \label{fig_fig1}
\end{figure}

{\it Electric quadrupolar current.}---To reveal the geometric origin of NDD, we first consider its static counterpart, i.e. a current driven by a spatially varying but static electric field in metals.  As $\bm\nabla \times \bm E = 0$, the electric field must be quadrupolar as shown in Fig.~\ref{fig_fig1}.  We will adopt the semiclassical transport theory since the band geometry enters naturally in the semiclassical equations of motion.  Below we sketch the derivation and leave the details in the Supplementary Materials~\cite{suppl}.  Our starting point is the equations of motion in crystals under slowly varying electric field
\begin{align}\label{eq_rdot}
\dot{\bm r}&=\frac{1}{\hbar}\bm \partial_{\bm k} \tilde{\varepsilon}_m-\dot{\bm k}\times \bm \Omega_{m}-\bm \Omega_{\bm k\bm r,m}\cdot\dot{\bm r}\,,\\
\hbar\dot{\bm k}&=-e\bm E+\bm \Omega_{\bm r\bm k,m}\cdot\hbar\dot{\bm k}\,,
\end{align}
where $\bm \Omega_m=-2{\rm Im}\langle \bm \partial_{\bm k}u_m|\times|\bm \partial_{\bm k}u_m\rangle$ is the momentum-space Berry curvature with $|u_m\rangle$ being the periodic part of Bloch functions in the $m$-th band, and $(\Omega_{\bm k\bm r,m})_{ij}=-2\IM\langle \partial_{k_i}u_m|\partial_{r_j}u_m\rangle$ is the mixed Berry curvature with $(\Omega_{\bm k\bm r,m})_{ij}=-(\Omega_{\bm r\bm k,m})_{ji}$. The appearance of $\bm \Omega_{\bm k\bm r,m}$ is due to the modification of the Bloch function by the inhomogeneous electric field.

The quantity of interest here is the band energy given by $\tilde{\varepsilon}_m=\varepsilon_m+\delta \varepsilon$, where $\varepsilon_m$ is the unperturbed band energy and $\delta \varepsilon=\frac{1}{2}e(\partial_iE_j)g_{ij,m}$ is the correction to $\varepsilon_m$~\cite{suppl}.  Here $g_{ij,m}$ is the Fubini-Study quantum metric~\cite{Provost1980,Resta2011}
\begin{equation}\label{eq_gij}
g_{ij,m} = \RE\bracket{\partial_{k_i} u_m|\partial_{k_j}u_m} - A_{i,m} A_{j,m} \;,
\end{equation}
where $A_{i,n} = \bracket{u_n|i\partial_{k_i}u_n}$ is the intraband Berry connection.  Geometrically, $g_{ij,m}$ measures the distance between neighboring Bloch states~\cite{Provost1980,Resta2011}. As the quantum metric enters in the energy correction by coupling to $\partial_iE_j$, it can be viewed as the electric quadrupole of Bloch states, consistent with the definition of the electric quadrupole in electromagnetic theory.

Since the system considered here breaks both time-reversal and inversion symmetry, a linear magnetoelectric coupling is allowed.  Consequently, a spatially varying electric field can induce a spatially varying magnetization, which gives rise to a magnetization current.  It is well established that the magnetization current should be discounted and the transport current reads (Eq.~(4) of Ref.~\cite{Xiao2006})
\begin{align}\label{eq_trans}
\bm J^\text{tr}&=-e\sum_m\int \frac{d\bm k}{8\pi^3}D\dot{\bm r}f_m\notag\\
&-\bm \nabla_{\bm r}\times \frac{e}{\hbar}\sum_m\int \frac{d\bm k}{8\pi^3}k_BT\bm \Omega_m \log\left(1+e^{\frac{\mu-\varepsilon_m}{k_BT}}\right)\,,
\end{align}
where $f_m$ is the Fermi function, and $D=1 + \TR\Omega_{\bm k\bm r,m}$ is modified density of states~\cite{Xiao2005}.

To evaluate Eq.~\eqref{eq_trans} at the order $\bm \partial\bm E$, we also need the response of the Berry curvature to the electric field, which reads $\bm \Omega_m^\prime=2e\bm \nabla_{\bm k}\times {\rm Re}\sum_{n\neq m}\bm A_{mn}(\bm A_{nm} \cdot \bm E) /\omega_{mn}$~\cite{Gao2014}, where $\omega_{mn}=\varepsilon_m-\varepsilon_n$ and $\bm A_{mn}=\bracket{u_m|i\bm \partial_{\bm k}u_n}$ is the interband Berry connection.

By plugging Eq.~\eqref{eq_rdot} and $\bm \Omega_m^\prime$ into Eq.~\eqref{eq_trans}, we can obtain the transport current. We find that only the term containing $\bm \partial_{\bm k}\delta \varepsilon$ remains and all other contributions cancel. We leave the detail in the Supplemental Materials~\cite{suppl}. The final result reads
\begin{equation} \label{eq_qcurrent}
J_x^\text{tr}=-2e\int \frac{d\bm k}{8\pi^3} f_m \partial_{k_x}\delta\varepsilon=\gamma_{xxz}\partial_zE_x \;,
\end{equation}
where
\begin{align}\label{eq_gamma}
\gamma_{jik}&=\frac{e^2}{\hbar}\sum_m\int \frac{d\bm k}{8\pi^3}G_{ijk,m}f_m^\prime\,,
\end{align}
Here $G_{ijk,m}=v_{i,m}g_{jk,m}$ with $v_{i,m}$ being the band velocity is the quantum metric dipole, defined similarly to the Berry curvature dipole~\cite{Inti2015}. The factor $2$ in Eq.~\eqref{eq_qcurrent} appears because in the static case $\partial_zE_x=\partial_xE_z$.  $f_m^\prime=\partial f_m/\partial \varepsilon$. Since the quantum metric is the Bloch state quadrupole, $G_{ijk,m}$ can also be viewed as the momentum space dipole of the electric quadrupole.  We emphasize that Eq.~\eqref{eq_qcurrent} represents the leading order transport current responsible for NDD at low frequency (as discussed later). Its connection to quantum metric dipole given in Eq.~\eqref{eq_gamma} is valid for any band structure with arbitrary number of bands.

Equation~\eqref{eq_qcurrent} is an intrinsic current independent of the transport relaxation time.
 It is also a Fermi surface effect and is hence important in metals and semiconductors. Since this current is in response to the variation of the electric field not the electric field itself, it persists even if the net electric field across the whole sample is zero.

{\it NDD at finite frequencies.}---We now reveal the geometric origin of NDD at finite frequencies. For this purpose, we need to express $\Delta n$ in Eq.~\eqref{eq_deltan} in terms of Bloch functions. We start from the standard Kubo formula of the optical conductivity
\begin{equation}\label{eq_opt}
\sigma_{ij}(\omega,\bm q)=-\frac{e^2}{i\omega}\sum_{m,n}\int \frac{d\bm k}{(2\pi)^3}\frac{(f_{m\bm k-\bm q}-f_{n \bm k}) M_{ij}}{\varepsilon_{m\bm k-\bm q}-\varepsilon_{n\bm k}+\hbar\omega+i\eta}\,,
\end{equation}
where $M_{ij}=\langle u_{m\bm k-\bm q}|\hat{v}_i|u_{n\bm k}\rangle\langle u_{n\bm k}|\hat{v}_j|u_{m\bm k-\bm q}\rangle$ and $\hat{\bm v}$ is the velocity operator.  It is obvious that the photon wave vector $\bm q$ shifts the crystal momentum and connects neighboring Bloch states in momentum space which are linked by the quantum metric.  As such, the quantum metric should appear naturally in NDD.

To explicitly demonstrate the role of quantum metric, we expand $\sigma_{xx}(\omega,\bm q)$ in Eq.~\eqref{eq_opt} to the linear order of $\bm q$. The general expression is presented in the Supplemental Materials~\cite{suppl}. The result can be simplified by considering the optical transition between two bands.  In this case we obtain~\cite{suppl}
\begin{equation}\label{eq_ndd}
\begin{split}
\Delta n &=-\frac{e^2c\mu_0}{2\hbar} \sum_{m,n=c,v}\int \frac{d\bm k}{8\pi^3}\Bigl[A (G_{zxx,m}f_m^\prime+G_{zxx,n}f_n^\prime) \\
&\qquad+B (G_{zxx,m}+G_{zxx,n}) +C (G_{xxz,m}+G_{xxz,n})\Bigr]\,.
\end{split}
\end{equation}
In Eq.~\eqref{eq_ndd} we have explicitly separated the geometric contribution from the spectral contribution.
The coefficients $A$, $B$ and $C$ only depend on the spectrum:
\begin{align}
A&=\frac{\omega_{mn}^2}{\omega_{mn}^2-\omega^2}+i\pi \omega \delta(\omega+\omega_{mn})\,,\notag\\
B&=-\frac{2\Delta f_{mn} \omega_{mn}^3}{(\omega_{mn}^2-\omega^2)^2}+i\pi\frac{\Delta f_{mn}\omega_{mn}^2}{\omega}\frac{d}{d\omega}\delta(\omega+\omega_{mn})\,,\notag\\
C&=\frac{2\Delta f_{mn}\omega_{mn}}{\omega_{mn}^2-\omega^2}-2i\pi\Delta f_{mn}\delta(\omega+\omega_{mn})\notag\,,
\end{align}
where $\Delta f_{mn}=f_m-f_n$.  The indices $m$ and $n$ run between the conduction~($c$) and valence~($v$) band.  We note that in insulators the expression for $\RE\Delta n$ is consistent with $\IM\sigma_{xxz}$ in Ref.~\cite{Malashevich2010}.

The quantum metric dipole $G$ fully conforms to the symmetry requirement for NDD. $g_{jk}$ is even under time reversal~($T$) and inversion~($I$) operations. Due to the appearance of the velocity, $G_{ijk}$ is odd under $T$ or $I$ and even under combined $TI$. Moreover, both $G_{xxz}$ and $G_{zxx}$ are odd under mirror-$z$ symmetry. Therefore, $\Delta n$ will vanish identically if $T$ or $I$ or mirror-$z$ symmetry is present.

Armed with the insight that NDD has its geometrical origin in the quantum metric dipole, next we exam the peak structure of $\Delta n$ as a function of $\omega$.  We will focus on $\IM\Delta n$ which can be measured by light absorbance.  In Eq.~\eqref{eq_ndd}, the imaginary parts of the first and third term contain delta functions and hence experience a peak when the averaged quantum metric dipole reaches a maxima, e.g., when the joint density of state experiences the Van Hove singularity.  However, the second term in Eq.~\eqref{eq_ndd} also contributes a peak of a different origin.  It contains the derivative of the delta function and can be rewritten as
\begin{equation}\label{eq_newpeak}
\frac{i\pi c\mu_0}{2\omega}\left.\frac{d}{d\lambda} \langle G_{zxx,m}+G_{zxx,n}\rangle\right |_{\lambda=-\omega}\,,
\end{equation}
where $\langle x\rangle(\lambda)=\sum_{m,n=c,v}\int (d\bm k/8\pi^3)x\Delta f_{mn}\omega_{mn}^2 \delta(\omega_{mn}-\lambda)$ is the average of $x$ for a fixed parameter $\lambda$. Equation~\eqref{eq_newpeak} shows that $\IM\Delta n$ will exhibit a peak when the averaged quantum metric dipole has the steepest slope. Due to the Kramers-Kronig relation, $\RE\Delta n$ should also have a peak at the same frequency.

\begin{figure}[t]
  \includegraphics[width=0.6\columnwidth]{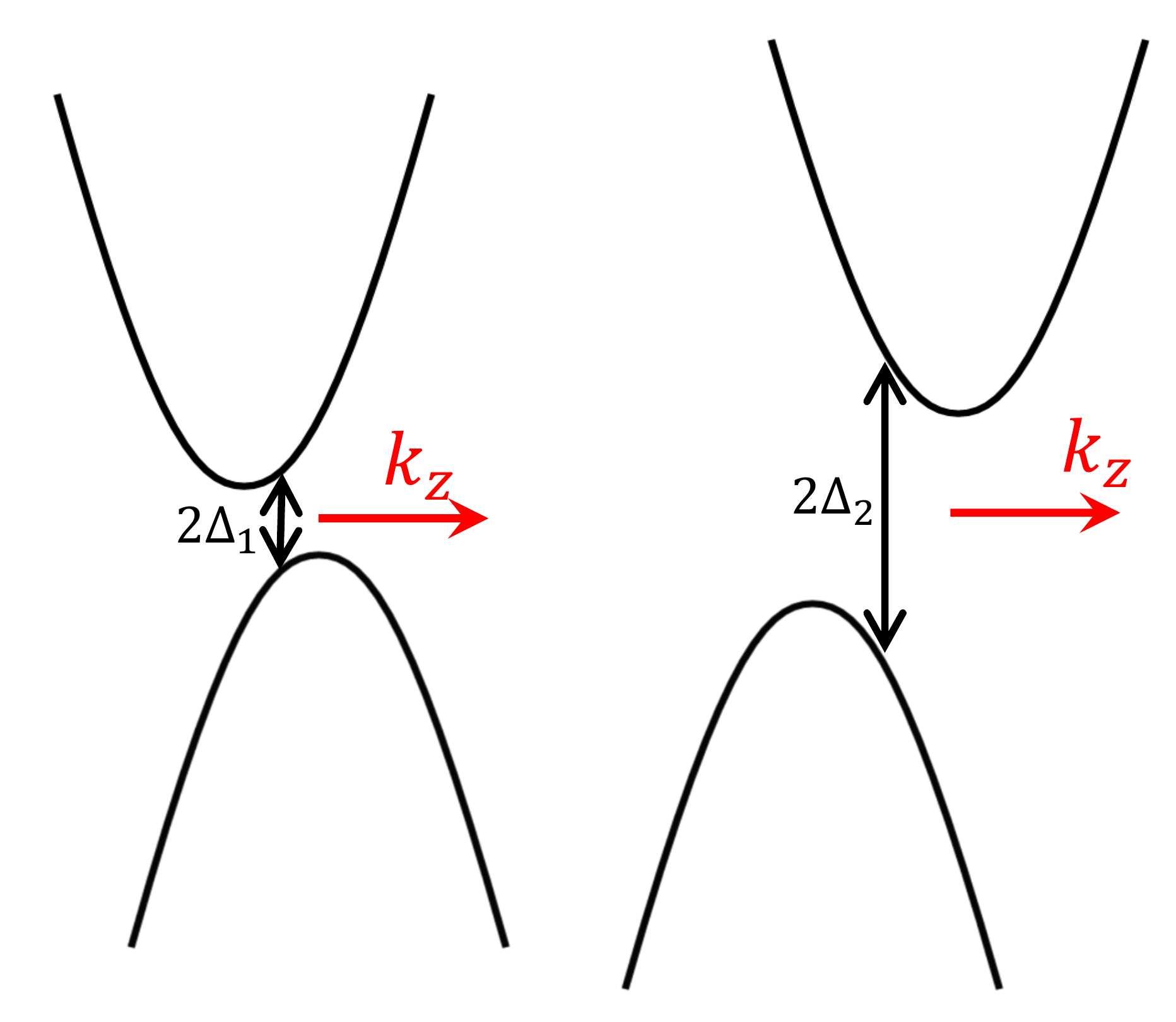}
  \caption{Energy spectrum near each valley. $v^\prime$ causes the conduction band bottom and the valence band top to shift oppositely along $k_z$, creating indirect band gaps. The energy difference at $k_z=0$ yields the band gap parameter $\Delta_i$.}\label{fig_fig2}
\end{figure}

{\it Low energy model.}---To demonstrate the geometrical features of NDD, we consider the following Hamilontian
\begin{equation}\label{eq_lowenergy}
\hat{H}=v_i^\prime k_z+v\tau_ik_x\sigma_x+ vk_y\sigma_y+ \left(\Delta_i+\frac{k_z^2}{2m_i}\right)\sigma_z\,,
\end{equation}
where $i$ is the valley index, and for each valley $\tau_i=\pm 1$, $2\Delta_i$ is the band gap, $v$ is the in-plane Fermi velocity, and $m_i>0$ is the effective mass along the $z$-direction.  The first term with $v_i^\prime$ introduces an anisotropy in the $z$-direction and hence effectively breaks the mirror-$z$ symmetry for each valley.  The energy spectrum around each valley is shown in Fig.~\ref{fig_fig2}. For fixed $k_z$, the corresponding 2D Hamiltonian is a gapped Dirac model that can be realized in layered two-dimensional magnets such as MnP$X_3$ ($X=$ S, Se) monolayers~\cite{Li2013,Sivadas2016}. The $k_z$ dependence can be introduced by stacking these 2D magnets along the $z$-direction and the $v^\prime$ term will appear when the stacking order breaks the mirror-$z$ symmetry.

\begin{figure}[t]
  \includegraphics[width=0.8\columnwidth]{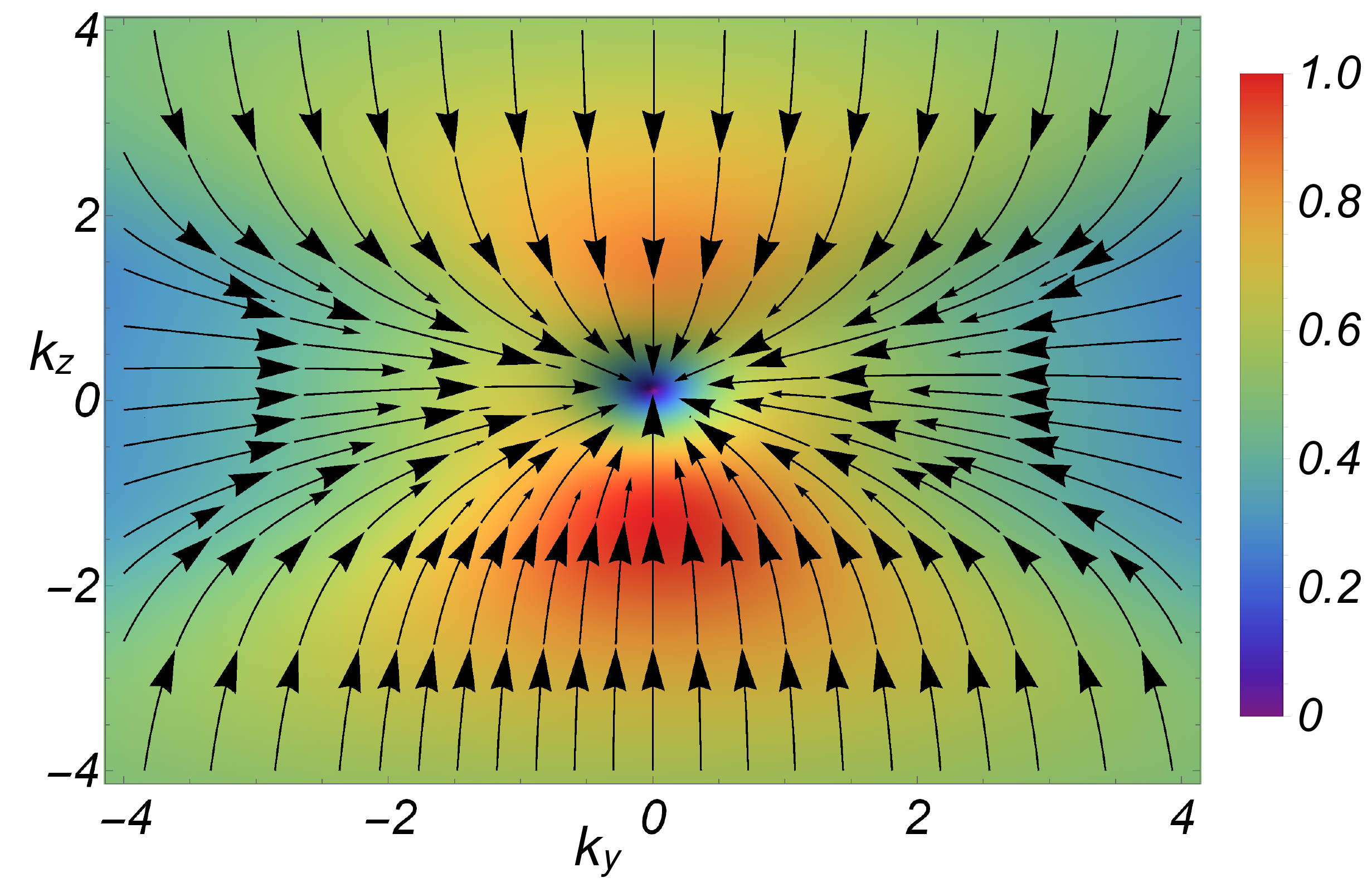}
  \caption{Sketch of the formation of a net quantum metric dipole. Streamline arrows and different colors show the direction and magnitude of the quantum metric dipole field, respectively. Although arrows are symmetric, the color map clearly has a dipolar structure, generating a net quantum metric dipole.}\label{fig_dgt}
\end{figure}

Since the valleys contribute additively, we shall limit our discussion to just one valley.  The low energy model has a net quantum metric dipole in each valley. For illustration purpose, we consider the dipoles of $g_{xx}$ for the valence band, which form a vector $(G_{xxx},G_{yxx},G_{zxx})$. After integrating this vector over $k_x$, we find that $G_{xxx}$ vanishes and the remaining components form a vector in the $k_y$-$k_z$ plane. In Fig.~\ref{fig_dgt} we sketch such vector field using streamline arrows and color map to represent its direction and magnitude, respectively. We observe that the direction distribution is symmetric. However, the magnitude distribution is only symmetric about $k_y=0$ axis. About $k_z=0$ axis, it has a clear dipolar structure with a stronger hot spot in the lower-half plane due to $v'_i$.

We now quantitatively demonstrate the influence of the quantum metric dipole on $\Delta n$ for a two-valley system.  For each valley, we calculate the averaged quantum metric dipole $\langle G\rangle=\langle G_{zxx,m}+G_{zxx,n}\rangle$ with the chemical potential in the band gap. We find that $\langle G\rangle= \Delta^2 S(\omega/\Delta) \sqrt{m{v^\prime}^2/2\Delta}\sqrt{\omega/(2\Delta) -1}$ with $S(\omega/\Delta)$ being a dimensionless structure factor~\cite{suppl}.  Clearly, the magnitude of $\langle G\rangle$ is proportional to the strength of the mirror-$z$ symmetry breaking term, $v^\prime$. Moreover, it has the steepest slope at $\omega=2\Delta$. Consequently, according to Eq.~\eqref{eq_newpeak}, $\Delta n$ has a term proportional to $1/\sqrt{\omega/(2\Delta)-1}$ and hence has a peak at $\omega=2\Delta$ (the band edge).

\begin{figure}[t]
  \includegraphics[width=\columnwidth]{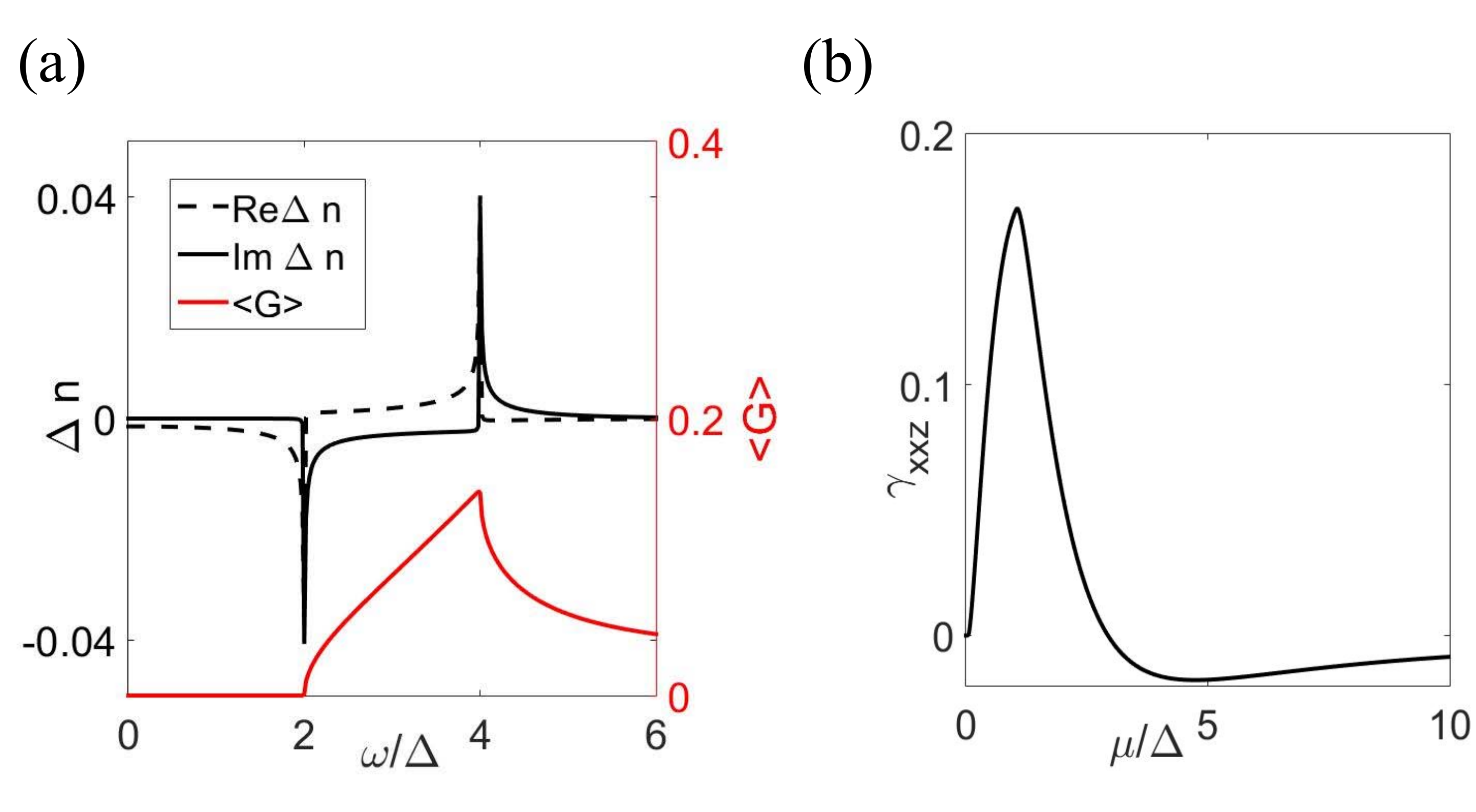}
  \caption{$\Delta n$~(Panel a) and the electric quadrupolar current~(Panel b) in the low-energy model. The parameters are chosen as follows: $\Delta_1=\Delta$, $\Delta_2=2\Delta$, $v_1^\prime=-v_2^\prime=v^\prime$, $\frac{1}{2}m(v^\prime)^2=0.95\Delta$. The system has a small global band gap $0.1\Delta$. In Panel a, $\langle G\rangle$ is in units of $\Delta^2$. In Panel b, $\gamma_{xxz}$ is in units of $e^2/(16\pi^2\hbar)$.}\label{fig_fig4}
\end{figure}

In Fig.~\ref{fig_fig4}a we plot $\Delta n$ and $G$ as a function of $\omega$ with the Fermi energy lies in the band gap. Clearly there are two peaks. This is because according to above analysis, for each valley, the slope of $\langle G\rangle$ has a maxima at $\omega=2\Delta_i$. Since $\Delta_1\neq \Delta_2$, there will be two peaks corresponding to the optical transition from each valley.  This peak structure remains in the metallic case when the chemical potential falls inside the band instead of the band gap~\cite{suppl}, confirming its generality.  Moreover, if the model has $T$ or $I$ symmetry, then $\Delta_2=\Delta_1$ and $v_1^\prime=-v_2^\prime$. The two peaks will appear at the same position with opposite values. Therefore, they will cancel each other and $\Delta n$ vanishes.

This low-energy model also allows the static electric quadrupolar current. In Fig.~\ref{fig_fig4}b, we calculate the relevant conductivity $\gamma_{xxz}$ at zero temperature. We find that $\gamma_{xxz}=0$ when $\mu$ lies in the global band gap.  As $\mu$ reaches the lower band bottom but still falls inside the gap of the other valley, $\gamma_{xxz}$ increases with the density of states.  When $\mu$ further increases, the other valley will also contribute but with opposite signs.  Hence $\gamma_{xxz}$ starts decreasing. When $\mu$ is large enough, $\gamma_{xxz}$ approaches $0$, as the quantum metric dipole decays faster than the increasing density of states.

{\it Connection to the DC transport current.}---We now demonstrate that the DC transport current in Eq.~\eqref{eq_qcurrent} can be obtained from the linear response result in Eq.~\eqref{eq_opt} by taking proper limit and then discounting local currents.  We first note that in the static case, $\partial_zE_x=\partial_xE_z$. Therefore, to recover our DC result, we need both the $xxz$- and $xzx$-components of the conductivity tensor, i.e., $J_x=\sigma_{xxz}\partial_zE_x+\sigma_{xzx}\partial_xE_z=(\sigma_{xxz}+\sigma_{xzx})\partial_zE_x$.  We focus on the dirty metal case with a non-vanishing $\eta$, in which the intrinsic current in Eq.~\eqref{eq_qcurrent} dominates over the extrinsic ones.  In this case, the static limit ($\omega\rightarrow 0$ and then $q\rightarrow 0$) and the uniform limit~($q\rightarrow 0$ and then $\omega\rightarrow 0$) yield the same intrinsic contribution~\cite{suppl},
\begin{align}\label{eq_jxxz}
{\rm Im}(\sigma_{xxz}+\sigma_{xzx})=-\alpha_{xy}+\frac{3}{2}\gamma_{xxz}+\frac{1}{2}\gamma_{xzx}\,,
\end{align}
where $\alpha_{xy}=P_x/B_y=M_y/E_x$ is the magnetoelectric coupling coefficient with $\bm P$ and $\bm M$ being the polarization and magnetization, respectively.  We take the imaginary part because $\bm \partial_{\bm r}=i\bm q$.

The current calculated from the Kubo formula~\eqref{eq_opt} is the total current, which, according to the Maxwell equation, can be written as
\begin{equation}
\bm J_\text{tot}=\frac{\partial \bm P}{\partial t}+\bm \nabla\times \bm M+\partial_j \dot{Q}_{ij} \hat{e}_i\,,
\end{equation}
where $Q_{ij}$ is the induced electric quadrupole.  In a transport experiment, the current being measured is the averaged current density over the whole sample~\cite{Cooper1997}.  After such a spatial average~\cite{suppl}, contributions due to both the magnetization, corresponding to $-\alpha_{xy}$, and the induced electric quadrupole moment, corresponding to $(\gamma_{xxz}+\gamma_{xzx})/2$, drop out from Eq.~\eqref{eq_jxxz}, and we are left with $\gamma_{xxz}$.  We thus recover the semiclassical DC result from the linear response theory.

In summary, we have identied the quantum metric dipole as the geometric origin of NDD. The static counterpart of NDD is a current driven by a quadrupolar electric field, obtained by integrating the quantum metric dipole over the Fermi surface.  Moreover, the steepest slope of the averaged quantum metric dipole yields a peak in NDD, which could be useful for band structure engineering of NDD.

\begin{acknowledgments}
We acknowledge useful discussions with Kenneth Burch, Nuh Gedik, Ajit Srivastava, Haidan Wen, Xiaodong Xu and Qi Zhang.  This work is supported by the Department of Energy, Basic Energy Sciences, Grant No.~DE-SC0012509.
\end{acknowledgments}

{\it Note added:}. The DC transport current in Eq.~\eqref{eq_qcurrent} has also been obtained independently in Ref.~\cite{lapa2019}.

\end{document}